\documentclass[universe,editorial,accept,pdftex,moreauthors]{Definitions/mdpi}

\firstpage{1} 
\pubvolume{10}
\issuenum{1}
\articlenumber{124}
\pubyear{2024}
\copyrightyear{2024}
\datereceived{28 February 2024 } 
\dateaccepted{28 February 2024  } 
\datepublished{5 March 2024 } 
\hreflink{https://\linebreak doi.org/10.3390/universe10030124} 
\doinum{10.3390/universe10030124}
\pdfoutput=1 

\usepackage{epigraph}

\Title{Top Quark at the New Physics Frontier}

\TitleCitation{Top Quark at the New Physics Frontier}

\Author{Efe Yazgan $^{1,}${*}$^{,\dagger}$\orcidA{} and Pedro Silva $^{2}$\orcidA{}}

\AuthorNames{Efe Yazgan and Pedro Silva}

\AuthorCitation{Yazgan, E.; Silva, P.}

\address{%
$^{1}$ \quad Department of Physics, National Taiwan University, Taipei 10617, Taiwan\\
$^{2}$ \quad  CERN, 1 Esplanade des Particules, EP-CMG-DS Bat 40 3-A32 E24410,  1211 Geneva 23, Switzerland; pedro.silva@cern.ch}

\corres{Correspondence: efe.yazgan@cern.ch}

\firstnote{Current address: CERN, 1 Esplanade des Particules, EP-UCM Bat 653/1-005 C00210, 1211 Geneva, Switzerland.} 

\begin{document}

\epigraph{Whaat? You work on the top quark? Stiill?}{\textit{Jack Steinberger to E.Y. \\ in Moriond QCD, La Thuile, March 2013}}

\vspace{6pt} 
This Special Issue on \href{https://www.mdpi.com/journal/universe/special_issues/Top_Quark_NPF}{''Top Quark at the New Physics Frontier''} is devoted to the most massive fundamental elementary particle known, the top quark. 
The aim is to provide a comprehensive review of the current status and prospects of top quark physics at the Large Hadron Collider (LHC) and future colliders. 
We included articles that emphasize where the present understanding is incomplete and suggest new directions for research in this area. 
We trust that it will benefit both those seeking to learn and those seeking to review recent developments in top quark physics.

The search for top quarks began half a century ago, with the prediction of the existence of the top quark through the six-quark model by Kobayashi and Maskawa~\cite{KobayashiMaskawa73}. In 1977, the $\Upsilon$ meson was discovered by the E288 experiment at Fermilab, providing the first evidence for the existence of three families of elementary particles~\cite{ref:bquark1,ref:bquark2}. 
The existence of the $b$ quark (the fifth quark) was inferred from interpreting the $\Upsilon$ meson as a bound state of a new heavy quark ($b$), and its anti-quark ($\overline{b}$) (see Ref.~\cite{ref:bquark2} and references therein). 
When interpreted in a quarkonium model, the measurements of the $\Upsilon$ decay width to $e^+e^-$ favored a b-quark charge of -1/3 instead of 2/3~\cite{bquark_charge1,bquark_charge2}. 
With this discovery, the down-type family of quarks ($d$, $s$, and $b$) was established, but only two up-type quarks were observed, the $u$ and the $c$ quark. 
The measurements of $Z\to b\bar{b}$, including the forward-backward asymmetry of b quarks and the $b\bar{b}$ cross section at $e^+e^-$ colliders, demonstrated the weak isopsin properties of the $b$ quark~\cite{SchaileZerwas1992}. With such properties, the $b$ quark surely had to be accompanied by  an upper isospin partner if the multiplet structure was present for the third quark family. 
Moreover, anomaly cancellation of the electroweak (EWK) gauge theory requires that the sum of the electric charges in a family is zero. 
This implies that another quark with a charge of +2/3 should exist.
An extra indication suggesting the existence of the top quark was the observation of fast oscillations of $B$--$\overline{B}$~\cite{argus,cleo}. 

Before the discovery of the top quark, using radiative corrections indirectly affecting the top quark and the measurements of EWK observables at a collision energy of $\sqrt{s}=100$ GeV, LEP1 predicted the top quark mass ($m_t$) to be $173^{+13}_{-10}$ GeV~\cite{ref:LEP1top}.
Eventually, the top quark was discovered in 1995 at the Fermilab Tevatron with $\mathcal{O}(10)$ signal events, independently by the CDF~\cite{ref:CDFtop} and D0~\cite{ref:D0top} experiments, using events from proton--antiproton ($p\overline{p}$) collisions at a center of mass energy of $\sqrt{s}=1.8$ TeV. 
Both experiments found a signal consistent with $t\overline{t}\to W^+b W^-\overline{b}$ events and inconsistent with the background prediction, and both were also able to reconstruct mass distributions with a clear peak. The measured mass values in each experiment with central values 176 GeV (with 7\% relative uncertainty) and 199 GeV (with 14\% relative uncertainty) were consistent with LEP1 predictions within uncertainties. 
This provided a critical test of the Standard Model (SM). 
Since then, many measurements of the top quark have been made both at the Tevatron and the LHC. Ten years before the discovery of the top quark, Lev B. Okun, outlined the conditions for {\it reliable experimental results}~\cite{okun}: 
\begin{quote}
{\it 
    The physics of elementary particles is done by people. It is characteristic of man to err
    ... Why then do physicists regard a multitude of phenomena as experimentally established, despite such mistakes?...
    How can it be guaranteed that these experiments are correct if so many incorrect results occurred in the past? The only guarantee is to accept a result as reliable only if it is obtained independently by several different groups employing different experimental methods. This condition is absolutely necessary but may not be sufficient, and does not provide a 100 percent guarantee. The 100 percent guarantee appears when the phenomenon recedes from the frontline of the science, when it is reproduced routinely, with the statistics of events exceeding by thousands or millions that with which the discovery was made, and when the quantities characterizing the phenomenon become known to an accuracy of several decimal places. Another way is not so much quantitative as it is qualitative: the search and discovery of a number of related phenomena that often follow the original discovery. } 
\end{quote}
It is fair to state that nearing 29 years after discovery, the top quark satisfies the ''absolute condition'' of a reliable result, because it has been experimentally established by five different experiments in different production modes using different methods and a variety of collision energies. 
Moreover, it is observed both in $p\overline{p}$ and $pp$ collisions.  Although it would be unfair to say that the top quark has receded from the frontline of science, its existence satisfies the ''100-percent guarantee'' proposed by Okun. Now, at the CERN LHC, top quark–antiquark pairs are routinely produced at a rate of about six per minute, enabling experiments to make detailed measurements of the properties of top quarks. Percent-level precision in several measurements has been possible thanks to the excellent performance of the Tevatron and LHC accelerators and experiments\footnote{Perhaps we can add one more condition for the top quark which is for now impossible to realize: that it becomes measurable in experiments independent of accelerators as well}. For the top quark, the ''qualitative condition'' that was listed by Okun is also met. The top quark and the Higgs boson modify the tree level SM processes through radiative corrections. Therefore, as was the case before the top quark discovery, the Higgs boson mass was predicted via an EWK fit. It was determined to be $94^{+25}_{-22}$ GeV, which turned out to be consistent with the  measured value of the Higgs boson mass within 1.3$\sigma$~\cite{ref:gfitter}. This test provides a high-precision consistency check of the SM. 

Significant progress has been made during the past years in improving  experimental measurements and computation techniques to achieve more 
accurate and precise quantum chromodynamics (QCD) calculations, EWK theory calculations, and Monte Carlo simulations. 
The reader will comprehend that a detailed review of all these developments would require extending this Special Issue to several journal issues. 
Therefore, our choice, as editors, was to emphasize recent LHC results and discuss future prospects in the field, and where relevant discuss the relation to cosmology (e.g. EWK baryogenesis, stability of the vacuum, dark matter, and axion-like particles) in the respective contributions. 

The top quark is an extraordinary elementary particle. It is the most massive elementary particle identified to date; not only does it have a privileged Yukawa coupling to the Higgs boson, it also has a mass that is significantly higher than that of the Higgs boson. 
At hadron colliders, top quarks are predominantly produced via QCD interactions. 
They are also produced ''alone'' through EWK interactions and are observed in single top quark channels. 
Owing to its large mass, the top quark decays before it can form a bound state, e.g. it can not form a $t\overline{t}$ meson ($toponium$). 
However, it can still be possible to observe some toponium effects in the phase space where the invariant mass of the $t\overline{t}$ pair is $2m_t\approx345$ GeV (e.g.~\cite{benjamin}). The top quark decays before hadronization, making the study of “bare” quark properties possible in experimental settings. Moreover, the spin-decorrelation timescale for $t\overline{t}$ pairs is larger than that of the hadronization time scale. This leaves the top and the anti-top quark spins correlated and allows them to stay $entangled$. This may allow for tests of the foundations of quantum theory at high energy scales.

Top quark physics simultaneously pushes the frontier of QCD, EWK, and flavor physics.  
Through top quark measurements in $t\overline{t}$ and single-top quark processes, the existence of many (new) physics phenomena is verified through top quark measurements. Some of these phenomena are discussed in this issue:  tests of charge-parity symmetry, lepton-flavor conservation, Lorentz-invariance violation and through that a precise test of special relativity, top quark Yukawa couplings via four-top quark production or via same-sign top quark plus a $c$ jet, triple top quark, or single-top quark plus $b$ jets to probe low-mass extra scalar particles, dark matter, axion-like particles, additional new particles such as color-octed vector $G$, neutral $Z'$ boson, or a charged $W'$ boson through $t\overline{t}$ asymmetries, and flavor-changing neutral currents of the top quark that connect the top quark with new scalar bosons. However, the analysis of the data collected by the first three LHC runs has revealed good agreement with the SM predictions. 
Currently, we do not have even a single direct or indirect indication of the existence of a new particle or interaction. Therefore, we are not in the same situation that we were in prior to the discoveries of the top quark and the Higgs boson. 

Now, without any direct or indirect indication of any new physics from the LHC, the scale of new physics is assumed to be above the TeV scale. Therefore, our focus has shifted to quantifying the effects of heavier hypothetical particles on our measurements at the LHC using the effective field theory (EFT) approach to identify dimension 6 operators that may affect our measurements, ensuring that kinematic distributions and cross sections align with observed data.
Along with the measurements of the top quark within the SM and direct model-dependent searches for new physics, results or calculations using the EFT approach are discussed in all contributions of this Special Issue, except one contribution that does not adopt this approach;  instead, this contribution promotes 
the general two-Higgs-doublet model (g2HDM), which offers two sets of dimension 4 operators to be investigated at the LHC and flavor experiments, specifically, new Yukawa and Higgs quartic couplings~\cite{george}. 
We embrace both approaches, however, we would have also welcomed a completely new revolutionary approach with no event generator, EFT, model-dependent search, or a Lagrangian.  
In any approach, precision measurements and open discussion are greatly needed, as emphasized by Robert B. Laughlin~\cite{laughlin}: 
\begin{quote}
{\it 
A measurement that cannot be done accurately can never be divorced from politics and must therefore generate mythologies. The more such shades of meaning there are, the less scientific the discussion becomes. Accurate measurement in this sense is scientific law, and a milieu in which accurate measurement is impossible is lawless. The need for precision, in turn, redoubles the need for that other great Greek tradition, open discussion for ideas and ruthless separation of meaningful things from meaningless ones. Precision alone does not guarantee good law...
}
\end{quote}

We can better understand what is meant by this quotation with the difficulties encountered in the interpretation of the precise measurements of the top quark mass (see e.g. Ref.~\cite{ref:hoang}). Using the measured values of the top quark and Higgs boson masses, one can say something about the stability of the EWK vacuum~\cite{ref:vacuumstab1,ref:vacuumstab2,ref:vacuumstab3}. Current values of the top quark and Higgs boson masses indicate that the EWK vacuum may be meta-stable and that the SM is consistent and could be valid up to the Planck scale~\cite{ref:Degrassi}. 
However, to be able to understand stability of the EWK vacuum, we need a few times better precision in top quark pole mass measurements. This requires an electron-positron collider or a much better understanding of the meaning of the Monte Carlo mass, especially, its relation to the so-called pole mass~\cite{ref:hoang}, or most probably the combination of both. 

In this Special Issue, you will find contributions covering all these topics, although often briefly, and in most cases without going into deep detail in the theoretical aspects. Where appropriate, the contributions include the prospects for top quark measurements and related new physics searches in experiments in future colliders such as HL-LHC, HE-LHC, FCC, ILC, CLIC, and CEPC.

\section*{List of Contributions}
\begin{enumerate}
    \item Hosseini, Y.; Najafabadi, M. M. Prospects for Probing Axionlike Particles at a Future Hadron Collider through Top Quark Production, {\em Universe}, {\bf 2022}, {\em 8}, 301. \url{https://doi.org/10.3390/universe8060301}.    
    \item Hou, G. W.-S. On Extra Top Yukawa Couplings of a Second Higgs Doublet, {\em Universe}, {\bf 2022}, {\em 8}, 475. \url{https://doi.org/10.3390/universe8090475}.
    \item Castro, N. F.; Skovpen, K. Flavour-Changing Neutral Scalar Interactions of the Top Quark, {\em Universe}, {\bf 2022}, {\em 8}, 609. \url{https://doi.org/10.3390/universe8110609}.
    \item Chwalek, T.; D\'eliot, F. Top Quark Asymmetries, {\em Universe}, {\bf 2022}, {\em 8}, 622. \url{https://doi.org/10.3390/universe8120622}.
    \item Blekman, F.; D\'eliot; F., Dutta; V., Usai, E. Four-top quark physics at the LHC,  {\em Universe}, {\bf 2022}, {\em 8}, 638. \url{https://doi.org/10.3390/universe8120638}.
    \item Behr, J. K.; Grohsjean, A. Dark Matter Searches with Top Quarks, {\em Universe}, {\bf 2023}, {\em 9}, 16. \url{https://doi.org/10.3390/universe9010016}.
    \item Thomas-Wilsker, J. Recent Cross-Section Measurements of Top-Quark Pair Production in Association with Gauge Bosons, {\em Universe}, {\bf 2023}, {\em 9}, 39. \url{https://doi.org/10.3390/universe9010039}.
    \item Chen, K. F.; Goldouzian, R. Tests of Charge–Parity Symmetry and Lepton Flavor Conservation in the Top Quark Sector, {\em Universe}, {\bf 2023}, {\em 9}, 62. \url{https://doi.org/10.3390/universe9020062}.
    \item D’Hondt, J.; Kim, T. J. Measurements of the Cross-Section for the $t\overline{t}$ Heavy-Flavor Production at the LHC, {\em Universe}, {\bf 2023}, {\em 9}, 242. \url{ https://doi.org/10.3390/universe9050242}.
    \item Andrea, J.; Chanon, N. Single-Top Quark Physics at the LHC: From Precision Measurements to Rare Processes and Top Quark Properties, {\em Universe}, {\bf 2023}, {\em 9}, 439. \url{https://doi.org/10.3390/universe9100439}.
    \item Pitt, M. Diffractive and Photon-Induced Production of Top Quark, {\em Universe}, {\bf 2023}, {\em 9}, 483. \url{https://doi.org/10.3390/universe9110483}.
    \item Jung, A. Properties of the Top Quark, {\em Universe}, {\bf 2024}, {\em 10}, 106. \url{https://doi.org/10.3390/universe10030106}.
\end{enumerate}

\funding{
E. Yazgan has received generous support from the Academic Summit
      Program of National Science and Technology Council of Taiwan (NSTC), Taiwan, as well as the National Taiwan University (NTU) President's Fund.
}

\acknowledgments{The editors are indebted to the authors and MDPI for their support and enthusiasm for the project, as well as for their cooperation. We also would like to gratefully acknowledge the CERN Scientific Information Service for their support in making this issue Open Access.}

\conflictsofinterest{The authors declare no conflicts of interest.} 

\begin{adjustwidth}{-\extralength}{0cm}

\reftitle{References}


\PublishersNote{}
\end{adjustwidth}
\end{document}